\documentstyle[12pt]{article}

\title{Vertex operator solutions to the discrete
KP-hierarchy\footnote{ The final version appeared in:
Comm. Math. Phys., {\bf 203}, 185--210 (1999)}}

\author{M. Adler\thanks{Department of Mathematics,
Brandeis University, Waltham, Mass 02454, USA. E-mail:
adler@math.brandeis.edu. The support of a National Science
Foundation grant \# DMS-9503246 is gratefully acknowledged.}~~~~~P.
van Moerbeke\thanks{Department of Mathematics, Universit\'e de
Louvain, 1348 Louvain-la-Neuve, Belgium and Brandeis University,
Waltham, Mass 02454, USA. E-mail: vanmoerbeke@geom.ucl.ac.be and
@math.brandeis.edu. The  support of a National Science Foundation
grant \# DMS-9503246, a Nato, a FNRS and a Francqui Foundation
grant is gratefully acknowledged.}}

\date{August 24, 1998}

\newcommand{\MAT}[1]{\left(\begin{array}{*#1c}}
\newcommand{\mat}{\end{array}\right)}
\newcommand{\qed}
{%
\mbox{}%
\nolinebreak%
\hfill%
\rule{2mm}{2mm}%
\medbreak%
\par%
}

\newcommand{\sumbis}[2]%
{%

\begin{array}[t]{c}
\sum\\
{\scriptstyle #1}\\
{\scriptstyle #2}
\end{array}

}

\newcommand{\rg}{\rightarrow}
\newcommand{\lrg}{\longrightarrow}
\newcommand{\Rg}{\Rightarrow}

\newcommand{\DR}{{\cal D}}

\newcommand{\BC}{{\Bbb C}}

\newcommand{\BZ}{{\Bbb Z}}

\newcommand{\iy}{\infty}
\newcommand{\pl}{\partial}
\newcommand{\al}{\alpha}
\newcommand{\proof}{\underline{\sl Proof}: }
\newcommand{\remark}{\underline{\sl Remark}: }

\newcommand{\HR}{{\cal H}}

\newcommand{\WR}{{\cal W}}
\newcommand{\FR}{{\cal F}}

\newcommand{\la}{\langle}
\newcommand{\ra}{\rangle}
\newcommand{\ga}{\gamma}

\newcommand{\dt}{\delta}

\newcommand{\vr}{\varepsilon}

\newcommand{\BR}{{\Bbb R}}
\newcommand{\lb}{\lambda}
\newcommand{\Lb}{\Lambda}

\def\span{\mathop{\rm span}}
\def\diag{\mathop{\rm diag}}

\def\be{\begin{equation}}
\def\ee{\end{equation}}
\def\bea{\begin{eqnarray}}
\def\eea{\end{eqnarray}}

\ifx\undefined\Bbb
        \let\Bbb\bf
\fi

\catcode`\@=11
\def\ps@X{\let\@mkboth\@gobbletwo
        \def\@oddhead{\tt Adler-van Moerbeke:%
        Discrete KP\hfil
        \date{August 24, 1998}\ \ \hfil\S\thesection,
p.\thepage
        }
        \def\@oddfoot{\rm\hfil\thepage\hfil}
        \let\@evenhead\@oddhead
        \let\@evenfoot\@oddfoot}
\catcode`@=12
\newtheorem{definition}{Definition}[
section]

\newtheorem{theorem}[definition]{Theorem}

\newtheorem{lemma}[definition]{Lemma}
\newtheorem{corollary}[definition]{Corollary}
\newtheorem{proposition}[definition]{Proposition}

\catcode`\@=11
\let\c@equation=\relax
\newcounter{equation}[
section]

\catcode`\@=12

\begin{document}
\maketitle

\tableofcontents

%

\vspace{1cm}

 {\em Vertex operators}, which are
disguised Darboux maps, transform solutions of the KP
equation into new ones. In this paper, we show that
the bi-infinite sequence obtained by Darboux
transforming an arbitrary KP solution recursively
forward and backwards, yields a solution to the {\em
discrete KP-hierarchy}. The latter is a KP hierarchy
where the continuous space $x$-variable gets replaced
by a discrete $n$-variable. The fact that these
sequences satisfy the discrete KP hierarchy is
tantamount to certain bilinear relations connecting
the consecutive KP solutions in the sequence. At the
Grassmannian level, these relations are equivalent to
a very simple fact, which is the nesting of the
associated infinite-dimensional planes (flag). The
discrete KP hierarchy can thus be viewed as a
container for an entire ensemble of vertex or Darboux
generated KP solutions.

It turns out that many new and old systems lead to such discrete
(semi-infinite) solutions, like sequences of soliton solutions,
with more and more solitons, sequences of Calogero-Moser systems,
having more and more particles, just to mention a few examples;
this is developped in \cite{AvM4}. In this paper, as an other
example, we show that the {\em $q$-KP hierarchy} maps, via a kind
of {\em Fourier transform}, into the discrete KP hierarchy,
enabling us to write down a very large class of solutions to the
$q$-KP hierarchy. This was also reported in a brief note with E.
Horozov\cite{AHV}.

\bigbreak

Given the shift operator $\Lb=(\dt_{i,j-1})_{i,j\in\BZ}$, consider
the Lie algebra
\be
\DR=\left\{\sum_{-\iy<i\ll\iy}a_i\Lb^i,a_i\mbox{\,\,diagonal
operators}\right\}=\DR_-+\DR_+
\ee
with the usual splitting $\DR=\DR_-+\DR_+$, into subalgebras
\be
\DR_+=\left\{\sum_{0\leq i\ll\iy}a_i\Lb^i\in\DR\right\},\DR_-=
\left\{\sum_{-\iy<i<0}a_i\Lb^i\in\DR\right\}.
\ee
The discrete
KP-hierarchy equations
\be
\frac{\pl L}{\pl t_n}=[(L^n)_+,L],\quad n=1,2,...
\ee
are deformations of an infinite matrix
\be
L=\sum_{-\iy<i\leq 0}a_i(t)\Lb^i+\Lb \in \DR,\quad\mbox{with
$t=(t_1,t_2,...)\in\BC^{\iy}$.}
\ee
If we represent $L$ as a dressing up of $\Lambda$ by a wave operator
$S\in I+\DR_-$
\be
L=S \Lb S^{-1}=W\Lb
W^{-1},\quad W=Se^{\sum_1^{\iy}t_i \Lb^i},
\ee
then the $L$-deformations are induced by $S$-deformations and
$W$-deformations:
\be
\frac{\pl S}{\pl t_n}=-(L^n)_-S,\quad\frac{\pl W}{\pl
t_n}=(L^n)_+W,\quad n=1,2,...;
\ee
In terms of vectors
\be
\chi(z)=(z^n)_{n\in\BZ},\quad\quad \chi^*(z)=\chi(z^{-1}),
\ee
such that
$ z\chi(z)=\Lb\chi(z),\quad
z\chi^*(z)=\Lb^{\top}\chi^*(z), $ let us define wave and adjoint wave
vectors $\Psi(t,z)$ and $\Psi^*(t,z)$
\be
\Psi(t,z)=W\chi(z)~\mbox{and}~
\Psi^*(t,z)=(W^{-1})^{\top}\chi^*(z).
\ee
We find, using (0.5), (0.8), (0.6), that
\bea
L\Psi(t,z)=z\Psi(t,z)  & & L^{\top}\Psi^*(t,z)=z\Psi^*(t,z),\nonumber\\
\frac{\pl\Psi}{\pl t_n}=(L^n)_+\Psi & & \frac{\pl\Psi^*}{\pl
t_n}=-((L^n)_+)^{\top}\Psi^*.
\eea

\begin{theorem}
If $L$ satisfies the Toda lattice, then the wave vectors $\Psi(t,z)$
and
$\Psi^*(t,z)$ can be expressed in terms of one sequence of
$\tau$-functions
$\tau(n,t):=
\tau_n(t_1,t_2,\dots),\quad n\in\BZ$,
to wit:
$$
\Psi(t,z)=\left(e^{\sum^{\iy}_1 t_iz^i}
\psi(t,z) \right)_{n\in\BZ}=\left(
        \frac{\tau_n(t-[z^{-1}])}{\tau_n(t)}
e^{\sum^{\iy}_1 t_iz^i} z^n
\right)_{n\in\BZ},
$$
\be
\Psi^*(t,z)=\left(e^{-\sum^{\iy}_1 t_iz^{i}}\psi^*(t,z)
\right)_{n\in\BZ}=\left(
\frac{\tau_{n+1}(t+[z^{-1}])}{\tau_{n+1}(t)}
e^{-\sum^{\iy}_1 t_iz^{i}}z^{-n}
\right)_{n\in\BZ},
\label{1.4}
\ee
satisfying the bilinear identity
\be
\oint_{z=\iy}\Psi_n(t,z)\Psi^*_m(t',z)\frac{dz}{2 \pi iz}=0
\ee
for all $n>m$. It follows that
$$\Psi=W\chi(z)=e^{\sum^{\iy}_1 t_iz^i} S
\chi(z),
$$
$$
\Psi^*=\left(W^{\top}\right)^{-1} \chi^{\ast}(z) = e^{-\sum^{\iy}_1
t_iz^i}(S^{-1})^{\top}\chi^{\ast}(z),
$$
with\footnote{In an expression, like $S=\sum a^{(n)} \Lambda^{n}$,
$a^{(n)}=\mbox{diag}(a^{(n)}_k)_{k \in \BZ}$ and $(\tilde
\Lambda a)_k=a_{k+1} \Lb^0 $.}
\be
S=\sum_0^{\iy}\frac{p_n(-\tilde\pl)\tau(t)}{\tau(t)}
\Lb^{-n}\quad\mbox{and}\quad
S^{-1}=\sum_0^{\iy}\Lambda^{-n}~\tilde \Lambda\left(\frac{
 p_n(\tilde\pl)\tau(t)} {\tau(t)}\right).
\ee
Then $L^k$ has the following expression in terms of
$\tau$-functions\footnote{where the $p_{\ell}$ are elementary Schur
polynomials and where $p_{\ell}(\tilde \pl)f \circ g$ refers to the
usual Hirota operation, to be defined in section 1.},
\be
L^k=\sum_{\ell=0}^{\iy}\mbox{diag}~\left(\frac{p_{\ell}(\tilde\pl)
\tau_{n+k-\ell+1}\circ\tau_n}
{\tau_{n+k-\ell+1} \tau_n}\right)_{n \in \BZ}\Lb^{k-\ell}
\ee
with the $\tau_n$'s satisfying
\be
\left(\frac{\pl}{\pl
t_k}-\sum^{\ell-1}_{r=0}(\ell-r)p_r(-\tilde\pl)p_{k-r}(\tilde\pl)\right)
\tau_{n}\circ\tau_{n-\ell}=0,~~\mbox{for}~\ell,k=1,2,3,...
\ee
and
$$
\left(\frac{1}{2}\frac{\pl^2}{\pl
t_1\pl
t_k}-p_{k+1}(\tilde\pl)\right)\tau_n\circ\tau_n=0,
~~\mbox{for}~k=1,2,3,...
$$
\end{theorem}

\noindent\remark Equation (0.14) reads
\bea
L^k&=&\Lb^k+\left(\frac{\pl}{\pl
t_1}\log\frac{\tau_{n+k}}{\tau_n}\right)_{n\in\BZ}
\Lb^{k-1}+...\nonumber\\
& &+\,\left(\frac{\pl}{\pl
t_k}\log\frac{\tau_{n+1}}{\tau_n}\right)_{n\in\BZ}\Lb^0+
\left(
\frac{\pl^2}{\pl t_1\pl t_k}\log\tau_n
\right)_{n\in\BZ}\Lb^{-1}+...\,,\nonumber\\
& &
\eea

\bigbreak

With each component of the wave vector $\Psi$, or, what is the same,
with each component of the $\tau$-vector, we associate a sequence of
infinite-dimensional planes in the Grassmannian
$Gr^{(n)}$
\bea
\WR_n&=&\mbox{ span}_{\BC}\left\{\left(\frac{\pl}{\pl t_1}
\right)^k\Psi_n(t,z),~~k=0,1,2,...\right\}\nonumber \\
&=&e^{\sum_1^{\iy}t_i z^i}\mbox{ span}_{\BC}\left\{\left(
\frac{\pl}{\pl
t_1}+z\right)^k\psi_n(t,z),~~k=0,1,2,...\right\}\nonumber\\
&=:& e^{\sum_1^{\iy}t_i z^i} \WR_n^t.
\eea

Note that the plane $z^{-n}\WR_n \in Gr^{(0)}$ has so-called virtual
genus zero, in the terminology of \cite{SW}; in particular,
this plane contains an element of order $1+O(z^{-1})$. Setting
$\{f,g\}=f'g-fg'$ for $'=\pl / \pl t_1$, we have the following
statement:

\bigbreak

\begin{theorem} The following six statements are
equivalent
\newline\noindent(i) The discrete KP-equations (0.3)
\newline\noindent(ii) $\Psi$ and $\Psi^*$, with the proper
asymptotic behaviour, given by (0.8),
satisfy the bilinear identities for all
$t,t' \in \BC^{\iy }$
\be
\oint_{z=\iy}\Psi_n(t,z) \Psi^*_m(t',z) \frac{dz}{2 \pi iz}=0,
~~~\mbox{ for all }~~n>m;
\label{7}\ee
\newline\noindent(iii)
the $\tau$-vector satisfies the following bilinear identities
for all $n>m$ and $t,t' \in \BC^{\iy }$:
\begin{equation}
\oint_{z=\iy}\tau_n(t-[z^{-1}])\tau_{m+1}(t'+[z^{-1}])
e^{\sum_1^{\iy}(t_i-t'_i)z^i}
z^{n-m-1}dz=0;
\label{8}
\end{equation}
\newline\noindent(iv) The components $\tau_n$ of a $\tau$-vector
correspond to a flag of planes in $Gr$,
\begin{equation}
... \supset \WR_{n-1}\supset
\WR_{n}\supset \WR_{n+1}\supset....
\end{equation}
\newline\noindent(v) A sequence of KP-$\tau$-functions $\tau_n$
satisfying the equations
\be\{ \tau_n
(t-[z^{-1}]),\tau_{n+1} (t)\} + z (\tau_n (t-[z^{-1}]) \tau_{n+1}
(t) - \tau_{n+1} (t-[z^{-1}])\tau_n (t)) = 0 \ee
\newline\noindent(vi) A sequence of KP-$\tau$-functions $\tau_n$
satisfying equations (0.14) for $\ell=1$, i.e.,
\be
\left(  \frac{\pl}{\pl t_k} - p_k(\tilde\pl) \right)\tau_{n+1}\circ
\tau_n  = 0 ~~\mbox{for}~k=2,3,...~\mbox{and}~n\in \BZ.
\ee
\end{theorem}

\vspace{0.5cm}

\noindent\remark The 2-Toda lattice, studied in \cite{UT}, amounts to
two coupled 1-Toda lattices or discrete KP-hierarchies, thus
introducing two sets of times
$t_n$'s and $s_n$'s. Actually, every 1-Toda lattice can naturally be
extended to a 2-Toda lattice; this is the content of Theorem 3.4.

\vspace{0.5cm}

\noindent{\bf How to construct discrete KP-solutions}. A wide class of
examples of discrete KP-solutions is given in section 4 by the following
construction, involving the simple vertex operators,
\be
X(t,z):=e^{\sum_1^{\iy}t_iz^i}e^{-\sum_1^{\iy}
\frac{z^{-i}}{i}\frac{\pl}{\pl t_i}},
\ee
which are disguised Darboux transformations acting on KP
$\tau$-functions. We now state:

\begin{theorem} Consider an arbitrary $\tau$-function for the KP
equation and a family of weights $...,\nu_{-1}(z)dz,\nu_0(z)dz,
\nu_1(z)dz,...$ on $\BR$. The infinite sequence of $\tau$-functions:
$\tau_0=\tau$ and, for $ n> 0$,
$$
\tau_{ n}: = \left(\int X (t,\lambda) \nu_{n-1}(\lb)d\lb...\int X
(t,\lambda)\nu_0(\lambda)d\lb\right)
\tau (t),
$$
$$
\tau_{-n} := \left(\int X (-t,\lambda) \nu_{-n}(\lb)d\lb...\int X
(-t,\lambda)\nu_{-1}(\lambda)d\lb\right)
\tau (t),
$$
form a discrete KP-$\tau$-vector, i.e., the bi-infinite matrix
\be
L=\sum_{\ell=0}^{\iy}\mbox{diag}~\left(\frac{p_{\ell}(\tilde\pl)
\tau_{n+2-\ell}\circ\tau_n}
{\tau_{n+2-\ell} \tau_n} \right)_{n \in \BZ}\Lb^{1-\ell}
\ee
satisfies the discrete KP hierarchy (0.3).
\end{theorem}

\noindent As an interesting special case of this situation, we study
in section 6 the {\em $q$-KP equation}.

\bigbreak
\noindent A wide variety of examples are captured by this
construction, like $q$-approximations to KP, discussed in section
5, but also soliton formulas, matrix integrals, certain integrals
leading to band matrices, the Calogero-Moser system and others,
discussed in \cite{AvM4}.

\noindent \underline{\em Remark}: A semi-infinite discrete
KP-hierarchy with $\tau_0(t)=1$ is equivalent to a bi-infinite discrete
KP-hierarchy with
$\tau_{-n}(t)=\tau_n(-t)$ and $\tau_0(t)=1$; this also amounts to
$\WR_{-n}=\WR^{\ast}_n$, with $\WR_{0}=\HR_+$. In such cases, one only
keeps the lower right hand corner of $L$, while the lower left hand
corner completely vanishes.

\section{The KP $\tau$-functions, Grassmannians and a
residue formula}

As is well known \cite{DJKM}, the bilinear identity
\be
\oint_{z=\iy}\Psi(t,z)\Psi^*(t,z)dz=0,
\ee
together with the asymptotics
\be
\Psi(t,z)=e^{\sum_1^{\iy}t_iz^i}\left(1+O\left(\frac{1}{z}\right)
\right),\Psi^*(t,z)=e^{-\sum_1^{\iy}t_iz^i}\left(1+O\left(\frac{1}{z}\right)
\right)
\ee
force $\Psi,\Psi^*$ to be expressible in terms of $\tau$-functions
$$
\Psi(t,z)=e^{\sum_1^{\iy}t_iz^i}\frac{\tau(t-[z^{-1}])}{\tau(t)},
\Psi^*(t,z)=e^{-\sum_1^{\iy}t_iz^i}\frac{\tau(t+[z^{-1}])}{\tau(t)};
$$
moreover the KP $\tau$-functions satisfy the differential Fay
identity\footnote{$\{  f,g\}:=\frac{\pl f}{\pl t_1} g-f \frac{\pl
g}{\pl t_1}$.}, for all $y,z\in\BC$, as shown in \cite{AvM1,vM}:
\bea
& &\{\tau(t-[y^{-1}]),\tau(t-[z^{-1}])\}\\
&
&\hspace{1cm}+\,(y-z)(\tau(t-[y^{-1}])\tau(t-[z^{-1}])-
\tau(t)\tau(t-[y^{-1}]
-[z^{-1}])=0.\nonumber
\eea
In fact this identity characterizes the $\tau$-function, as shown in
\cite{TT}.

From (1.1), it follows that
\bea
0&=&\oint\tau(t-a-[z^{-1}])\tau(t+a+[z^{-1}])e^{-2\sum_1^{\iy}
a_iz^i}\frac{dz}{2\pi i}\nonumber\\
& &=\sum^{\iy}_{k=1}a_k\left(\frac{\pl^2}{\pl t_1\pl
t_k}-2p_{k+1}(\tilde\pl)\right)\tau\circ\tau+O(a^2).
\eea
The Hirota notation used here is the following: Given a polynomial
$p\left(\frac{\pl}{\pl t_1},\frac{\pl}{\pl t_2},...\right)$ in
$\frac{\pl}{\pl t_i}$, define the symbol
\be
p\left(\frac{\pl}{\pl t_1},\frac{\pl}{\pl
t_2},...\right)f\circ g(t):=p\left(\frac{\pl}{\pl
u_1},\frac{\pl}{\pl u_2},...\right)f(t+u)g(t-u)\Biggl|_{u=0},
\ee
and
$$
\tilde\pl_t:=\left(\frac{\pl}{\pl t_1},\frac{1}{2}\frac{\pl}{\pl
t_2},\frac{1}{3}\frac{\pl}{\pl t_3},...\right).
$$

For future use, we state the following proposition shown in
\cite{AvM1}:

\begin{proposition}  Consider
$\tau$-functions
$\tau_1$  and $\tau_2$, the corresponding wave functions
\be
\Psi_j = e^{\sum_{i\geq 1} t_i z^i} {\tau_j (t-[z^{-1}])\over
\tau_j (t)}=e^{\sum_{i\geq 1} t_i z^i}\left(1+O(z^{-1})\right)
\ee
and the associated infinite-dimensional planes, as points in the
Grassmannian $Gr$,
$$\tilde \WR_i=\span\left\{\left(
\frac{\pl}{\pl t_1}\right)^k \Psi_i(t,z), \mbox{ for }
k=0,1,2,...\right\}~~\mbox{with}~~\tilde
\WR^t_i=\tilde \WR_i e^{-\sum_1^{\iy}t_kz^k};
$$
then the following statements are equivalent
\newline\noindent(i) $ z \tilde \WR_2 \subset \tilde \WR_1 $;
\vspace{0.2cm}\newline\noindent(ii) $ z \Psi_2 (t,z) =
\frac{\pl}{\pl t_1}
\Psi_1 (t,z) - \alpha \Psi_1 (t,z)$, for some function
$\alpha = \alpha (t)$;
\vspace{0.2cm}\newline\noindent(iii)\be\{ \tau_1
(t-[z^{-1}]),\tau_2 (t)\} + z (\tau_1 (t-[z^{-1}]) \tau_2 (t)
- \tau_2 (t-[z^{-1}])\tau_1 (t)) = 0 .\ee
\newline\noindent When (i), (ii) or (iii) holds, $\alpha(t)$ is
given by
\be
\alpha (t) = {\pl \over \pl t_1} \log {\tau_2 \over \tau_1}.
\ee
\end{proposition}

\proof To prove that (i) $\Rightarrow$ (ii), the inclusion
$z\tilde \WR_2 \subset \tilde \WR_1$, hence
$z\tilde \WR^t_2 \subset \tilde \WR^t_1$,  implies by (0.16) that
$$
z \psi_2 (t,z) = z (1+O(z^{-1})) \in \tilde \WR_1^t
$$
must be a linear combination\footnote{$\psi_i$ is the same as
$\Psi_i$, but without the exponential.}
\be
z \psi_2 = {\pl \psi_1\over
\pl x} + z \psi_1 -\alpha (t)  \psi_1,\mbox{ and thus }
z
\Psi_2 = {\pl \over \pl t_1} \Psi_1 - \alpha (t) \Psi_1.
\ee
The expression (1.8) for $\alpha (t)$ follows from equating the
$z^0$-coefficient in (ii), upon using the $\tau$-function
representation (1.6). To show that (ii) $\Rightarrow$ (i), note
that
$$
z \Psi_2 = {\pl \over \pl t_1} \Psi_1 - \alpha \Psi_1 \in \tilde\WR_1
$$
and taking $t_1$-derivatives, we have
$$
z \left( \frac{\pl}{\pl t_1}\right)^j \Psi_2 =  \left(
\frac{\pl}{\pl t_1}\right)^{j+1} \Psi_1 + \beta_1 \left(
\frac{\pl}{\pl t_1}\right)^j \Psi_1 + \cdots + \beta_{j+1}
\Psi_1,
$$
for some $\beta_1,\cdots,\beta_{j+1}$ depending on $t$ only;
this implies the inclusion (i). The equivalence (ii)
$\Longleftrightarrow$ (iii) follows from a straightforward
computation using the $\tau$-function representation (1.6) of (ii)
and the expression for $\alpha (t)$.\qed

\begin{lemma} The following integral along a clockwise circle in the
complex plane encompassing $z=\iy$ and $z=\al^{-1}$, can be evaluated as
follows
\begin{eqnarray*}
& &\oint_{z=\iy}
f(t+[\al]-[z^{-1}])g(t-[\al]+[z^{-1}])\frac{z^{m+1}}{(z-
\al^{-1})^2}\frac{dz}{2\pi iz}\\
&=&\al^{1-m}\sum^{\iy}_{k=1}\al^k\left(-\frac{\pl}{\pl
t_k}+\sum_{r=0}^{m-1}(m-r)p_r(-\tilde\pl)p_{k-r}(+\tilde\pl)\right)
f\circ
g.
\end{eqnarray*}
\end{lemma}

\proof By the residue theorem, the integral above is the sum of
residue at $z=\iy$ and at $z=\al^{-1}$:
\bea
&
&\oint_{z=\iy}f(t+[\al]-[z^{-1}])g(t-[\al]+[z^{-1}])\frac{z^{m+1}}{
(z-\al^{-1})^2}
\frac{dz}{2\pi iz}\nonumber\\
&=& \frac{1}{(m-1)!}\left(
\frac{d}{du}\right)^{m-1}f(t+[\al]-[u])g(t-[\al]+[u])
\frac{1}{(1-u\al^{-1})^2}\Biggl|_{u=0}\nonumber\\
& & \\
&
&-\frac{d}{dz}z^mf(t+[\al]-[z^{-1}])g(t-[\al]+[z^{-1}])\Biggl|_{z=\al
^{-1}}.
\eea
Evaluating each of the pieces requires a few steps.

\medbreak

{\bf Step 1.}
$$
\frac{1}{k!}\left(\frac{d}{du}\right)^kf(t+[\al]-[u])g(t-[\al]+
[u])\Biggl|_{u=0}=\sum^{\iy}_{\ell=0}\al^{\ell}p_k(-\tilde\pl)p_{\ell}
(\tilde\pl)f\circ g.
$$
At first note
\be
\left(\frac{d}{du}\right)^kF([u])\Biggl|_{u=0}=k!p_k(\tilde\pl_s)F(s)
\ee
and, by (1.5) and (1.12),
\bea
\frac{1}{k!}\left(\frac{d}{du}\right)^kf(t+[u])g(t-[u])
\Biggl|_{u=0}&=&p_k(\tilde\pl)f\circ g  \nonumber\\
&=&p_k(-\tilde\pl)g\circ f  \nonumber\\
&=& \sum_{i+j=k} p_i(-\tilde \pl)g . p_j(\tilde \pl)f.
\eea
Indeed
\medbreak

$\displaystyle{\frac{1}{k!}\left(\frac{d}{du}\right)^kf(t+[\al]-[u])
g(t-[\al]+[u])\Biggl|_{u=0}}$
\begin{eqnarray*}
&=&p_k(\tilde\pl_s)g(t-[\al]+s)f(t+[\al]-s)\Biggl|_{s=0},\quad
\mbox{using (1.12)}\\
&=&p_k(\tilde\pl_s)\sum^{
\iy}_{\ell=0}\al^{\ell}p_{\ell}(\tilde\pl_t)f(t-s)\circ
g(t+s)\Biggl|_{s=0},\quad\mbox{using (1.13)}\\
&=&\sum^{\iy}_{\ell=0}\al^{\ell}p_k(\tilde\pl_s)p_{\ell}(\tilde\pl_w)
f(t+w-s)g(t-w+s)\Biggl|_{s=w=0},\quad\mbox{expressing Hirota,}\\
&=&\sum^{\iy}_{\ell=0}\al^{\ell}p_k(\tilde\pl_s)p_{\ell}(-\tilde\pl_w)
f(t-w-s)g(t+w+s)\Biggl|_{s=w=0},\quad\mbox{flipping signs,}\\
&=&\sum^{\iy}_{\ell=0}\al^{\ell}p_k(\tilde\pl_v)p_{\ell}(-\tilde\pl_v)
f(t-v)g(t+v)\Biggl|_{v=0}\\
&=&\sum^{\iy}_{\ell=0}\al^{\ell}p_k(-\tilde\pl)p_{\ell}(\tilde\pl)
f\circ g,~~~~~~\mbox{using} (1.13).
\end{eqnarray*}

\medbreak

{\bf Step 2.} Residue at $\iy$.

Note
\be
\left(\frac{d}{du}\right)^{\ell}\left(\frac{1}{1-u\al^{-1}}\right)^2
\Biggl|_{u=0}=\left(\frac{d}{du}\right)^{\ell}
\sum^{\iy}_{i=1}i(u\al^{-1})^{i-1}\Biggl|_{u=0}=(\ell+1)!\al^{-\ell};
\ee
then we find
\medbreak

$\displaystyle{\frac{1}{(m-1)!}\left(\frac{d}{du}\right)^{m-1}f(t+
[\al]-[u])
g(t-[\al]+[u])\frac{1}{(1-u\al^{-1})^2}\Biggl|_{u=0}}$
\bea
&=&\frac{1}{(m-1)!}\sum^{m-1}_{r=0}\MAT{1}m-1\\r\mat
\left(\frac{d}{du}\right)^rf(t+[\al]-[u])g(t-[\al]+[u])
\nonumber\\
&
&\hspace{6cm}\left(\frac{d}{du}\right)^{m-1-r}\frac{1}{(1-u\al^{-1})^2}\Biggl|_{u=0}\nonumber\\
&=&\sum^{m-1}_{r=0}(m-r)\sum_{\ell=0}^{\iy}
\al^{\ell-m+r+1}p_r(-\tilde\pl)p_{\ell}(\tilde\pl)f\circ
g,\quad\mbox{using step 1 and (1.14)}\nonumber\\
&=&m\al^{1-m}f(t)g(t)+\al^{1-m}
\sum^{\iy}_{k=1}\al^k\sum^m_{r=0}(m-r)p_r(-\tilde\pl)
p_{k-r}(\tilde\pl)f\circ g,\quad\mbox{using $p_0=1$}.\nonumber\\
& &
\eea

\medbreak

{\bf Step 3.} Residue at $z=\al^{-1}$.

\medbreak

$\displaystyle{\frac{d}{dz}z^mf(t+
[\al]-[z^{-1}])
g(t-[\al]+[z^{-1}])\Biggl|_{z=\al^{-1}}}$
\bea
&=&-u^2\frac{d}{du}u^{-m}f(t+[\al]-[u])g(t-[\al]+[u])\Biggl|_{u=\al}\nonumber\\
&=&m\al^{-m+1}f(t)g(t)-\al^{2-m}\frac{d}{du}f(t+[\al]-[u])
g(t-[\al]+[u])\Biggl|_{u=\al}\nonumber\\
&=&m\al^{1-m}f(t)g(t)+
\sum^{\iy}_{k=1}\al^{1-m+k}\frac{\pl}{\pl
t_k}f\circ g,\quad\mbox{by explicit differentiation.}\nonumber\\
& &
\eea
Finally, putting step 2 and step 3 in (1.11) yields Lemma 1.2.\qed

\begin{lemma} The Hirota symbol acts as follows on functions
$f(t_1,t_2,...)$ and $g(t_1,t_2,...)$:
\begin{equation}
\frac{1}{fg}\frac{\pl^n}{\pl t_1...\pl t_n}f\circ g ~~=\mbox{ a
polynomial $P_n$ in}~\left\{
\begin{array}{l}
\frac{\pl^k}{\pl t_{i_1}...\pl t_{i_k}}\log \frac{f}{g}~~~\mbox{for $k$
odd}\\
\\
\frac{\pl^k}{\pl t_{i_1}...\pl t_{i_k}}\log fg~~~\mbox{for $k$
even}
 \end{array}
\right.
\end{equation}
over all subsets $\{i_1,...,i_k\} \subset \{1,...,n\}$. Upon
granting degree $1$ to each partial in $t_i$, the polynomial $P_n$ is
homogeneous of degree $n$.
\end{lemma}

\proof By induction, we assume the statement to be valid for an Hirota
symbol, involving $\ell$ partials, and we prove the statement for a
symbol involving $\ell +1$ partials:
$$
\frac{1}{fg}\frac{\pl}{\pl t_{\ell +1}}\frac{\pl^{\ell}}{\pl t_1...\pl
t_{\ell}}f(t)\circ  g(t)\hspace{7cm}
$$
\bea
&=&\frac{1}{fg}\frac{\pl}{\pl
u_{\ell+1}}f(t+u)g(t-u)\frac{\frac{\pl^{\ell}}{\pl t_1...\pl
t_{\ell}}f(t+u)\circ  g(t-u)}{f(t+u)g(t-u)}\Big|_{u=0}\nonumber\\
&=&\left( \frac{\pl}{\pl t_{\ell+1}}  \log \frac{f}{g}
\right)\frac{1}{fg}\frac{\pl^{\ell}}{\pl t_1...\pl
t_{\ell}}f(t+u)\circ  g(t-u) \nonumber\\
& & +\frac{\pl}{\pl u_{\ell+1}}P\left(
...,\frac{\pl^m}{\pl t_{i_1}...t_{i_m}}\log \frac{f(t+u)}{g(t-u)},...,
\frac{\pl^n}{\pl t_{j_1}...\pl t_{j_n}}\log f(t+u)g(t-u),...
\right)\Big|_{u=0}, \nonumber\\
\eea
where $m$ is odd and $n$ even. The result follows from the simple
computation:
\bea
\frac{\pl}{\pl u_{\ell+1}} \frac{\pl^m}{\pl t_{i_1}...\pl t_{i_m}}\log
\frac{f(t+u)}{g(t-u)}\Big|_{u=0}&=&\frac{\pl^{m+1}}{\pl
t_{i_1}...\pl t_{i_m}.\pl t_{\ell+1}}\log   f(t)g(t)\nonumber\\
\frac{\pl}{\pl u_{\ell+1}} \frac{\pl^n}{\pl t_{i_1}...\pl t_{i_n}}\log
f(t+u)g(t-u)\Big|_{u=0}&=&\frac{\pl^{n+1}}{\pl
t_{i_1}...\pl t_{i_n}.\pl t_{\ell+1}}\log   \frac{f(t)}{g(t)}\nonumber\\
\eea
\qed

\remark The induction formula (1.18) can be made into an explicit
formula for $P_n$, involving partitions of the set $\{1,2,...,n\}$.

\section{The existence of a $\tau$-vector and the discrete KP
bilinear identity}

Before proving Theorem 0.1, we shall need two lemmas, which
are analogues of basic lemmas in the theory of differential
operators. So the main purpose of this section is
threefold, namely, to prove the bilinear identities for the
wave and adjoint wave vectors, to prove the existence of a
$\tau$-vector and finally to give a closed form for
$L^k$.

\begin{lemma} For $z$-independent $U,~V \in \DR$,
the following matrix identities hold \footnote{
        $(A\otimes B)_{ij}=A_iB_j$ and remember
$\chi^*(z)=\chi(z^{-1})$. The contour in the integration below runs
clockwise about $\iy$; i.e., opposite to the usual orientation.}
\be
UV
=\oint_{z=\iy}U \chi(z)\otimes
V^\top\chi^*(z)\frac{dz}{2\pi iz},
\ee
\end{lemma}

\proof Set
$$
U=\sum_\al u_{\al}\Lb^\al\quad\hbox{and}\quad
V=\sum_\beta\Lb^\beta v_{\beta},
$$
where $u_{\al}$ and $v_{\al}$ are diagonal matrices.
To prove (2.1), it suffices to compare the
$(i,j)$-entries on each side. On the left side of (2.1), we have
\begin{eqnarray*}
\left(UV\right)_{ij}
&=&\Bigl(\sum_{\al,\beta}u_{\al}\Lb^{\al+\beta}
v_{\beta}\Bigr)_{ij}
\\
&=&\sum_{\al,\beta}u_{\al}(i)
(\Lb^{\al+\beta})_{ij}v_{\beta}(j)
\\
&=&\sum_{{\al,\beta}\atop{\al+\beta=j-i}}u_{\al}(i)
v_{\beta}(j).
\end{eqnarray*}
On the right side of (2.1), we have
\begin{eqnarray*}
\lefteqn{
        \oint_{z=\iy}\Bigl(U\chi(z)\Bigr)_i
        \Bigl(V^\top\chi(z^{-1})\Bigr)_{j}
        ~\frac{dz}{2\pi iz}
}\\
&=&\oint_{z=\iy}\Bigl(\sum_{\al}u_{\al}z^{\al}\chi(z)\Bigr)_i
\Bigl(\sum_{\beta}v_{\beta}z^{\beta}\chi(z^{-1})\Bigr)_{j}
\frac{dz}{2\pi iz}\\
&=&\oint_{z=\iy}
\sum_{\al,\beta}
u_{\al}(i)v_{\beta}(j)z^{\al+\beta+i-j}\frac{dz}{2\pi iz}\\
&=&
\sum_{{\al,\beta}\atop{\al+\beta=j-i}}
u_{\al}(i)v_{\beta}(j),
\end{eqnarray*}
establishing (2.1). \qed

\begin{lemma} For $W(t)$ a wave operator of the discrete
KP-hierarchy,
\be
W(t)W^{-1}(t')\in\DR_+,\quad\forall t,t'.
\ee
\end{lemma}

\proof Setting $h(t,t')=W(t)W^{-1}(t')$, compute from (0.6)
\be
\frac{\pl h}{\pl t_n}=(L^n(t))_+h,\quad\frac{\pl h}{\pl
t'_n}=-h(L^n(t'))_+,
\ee
since $h(t,t)=I\in\DR_+$, it follows that $h(t,t')$ evolves in
$\DR_+$.\qed

\medbreak

Consider the wave function, already defined in the introduction, and
the adjoint wave function:
\bea
\Psi(t,z)&=&W\chi(z)=e^{\sum_1^{\iy}t_iz^i}S\chi(z)=
e^{\sum
t_iz^i}\left(z^n+\sum_{i<n}s_i(n)z^i \right)_{n\in\BZ}
\nonumber\\
\Psi^*(t,z)&=&(W^{-1})^{\top}\chi^*(z)=e^{-\sum_1^{\iy}t_iz^i}
(S^{-1})^{\top} \chi^*(z)\nonumber\\
& &\quad\quad\quad =e^{-\sum
t_iz^i}\left(z^{-n}+\sum_{i<-n}s_i^{\ast}(n)z^i
\right)_{n\in\BZ}.
\eea

{\underline{\sl Proof of Theorem 0.1}: }

{\bf Step 1:} Setting
$$
U:=W(t)\quad\mbox{and}\quad V^{\top}:=(W^{-1}(t'))^{\top}
$$
in formula (2.1) of Lemma 2.1, and using formula (0.8) of $\Psi$ and
$\Psi^*$ in terms of $W$, one finds for all $t,t'\in\BC^{\iy}$,
\be
W(t)W(t')^{-1}=\oint_{z=\iy}\Psi(t,z)\otimes\Psi^*(t',z)\frac{dz}{2\pi
iz}.
\ee
But, according to Lemma 2.2, $W(t)W(t')^{-1}\in\DR_+$ and thus
(2.5) is upper-triangular, yielding
\be
\oint_{z=\iy}\Psi_n(t,z)\Psi^*_m(t',z)\frac{dz}{2\pi
iz}=0\quad\mbox{for all $n>m$}.
\ee
Defining
\begin{eqnarray*}
\Phi_n(t,z)&:=&z^{-n}\Psi_n(t,z)=e^{\sum
t_iz^i}(1+O(z^{-1}))\\
\Phi_n^*(t,z)&:=&z^{n-1} \Psi^*_{n-1}(t,z)=e^{-\sum
t_iz^i}(1+O(z^{-1})),
\end{eqnarray*}
upon using the asymptotics (0.8), we have, by setting $m=n-1$ in
(2.6)
$$
\oint_{z=\iy}\Phi_n(t,z)\Phi^*_n(t',z)dz=
\oint_{z=\iy}\Psi_n(t,z)\Psi^*_{n-1}(t',z)\frac{dz}{z}=0.
$$
From the KP-theory, there exists a $\tau$-function $\tau_n(t)$ for
each $n$, such that
$$
\Phi_n(t,z)=e^{\sum
t_iz^i}\frac{\tau_n(t-[z^{-1}])}{\tau_n(t)},\quad\Phi^*_n(t,z)=e^{-\sum
t_iz^i}\frac{\tau_n(t+[z^{-1}])}{\tau_n(t)},
$$
yielding the $\tau$-function representation (0.10) for $\Psi_n$ and
$\Psi^*_n$.\qed

\medbreak

{\bf Step 2:} The following holds for $n\in\BZ$:
\be
\left(\frac{1}{2}\frac{\pl^2}{\pl
t_1\pl
t_k}-p_{k+1}(\tilde\pl)\right)\tau_n\circ\tau_n=0,
~~\mbox{for}~k=1,2,3,...
\ee
\be
\left(\frac{\pl}{\pl
t_k}-\sum^{\ell-1}_{r=0}(\ell-r)p_r(-\tilde\pl)p_{k-r}(\tilde\pl)\right)
\tau_{n}\circ\tau_{n-\ell}=0,~~\mbox{for}~\ell,k=1,2,3,...
\ee
Indeed the bilinear identity (2.6), upon setting $m=n-\ell-1$, shifting
$t\mapsto t+[\al]$, $t'\mapsto t-[\al]$, using the
$\tau$-function representation (0.10) of $\Psi$ and $\Psi^*$, and lemma
1.2 with $m=\ell$, yield\footnote{
$e^{m\sum_1^{\iy}(\al z)^i/i}=(1-\al z)^{-m}$}
\begin{eqnarray*}
0&=&-\al^2\oint_{z=\iy}\Psi_n(t+[\al],z)\Psi^*_{n-\ell-1}
(t-[\al],z)\frac{dz}{2\pi iz}\tau_n(t+[\al])\tau_{n-\ell
}(t-[\al])\\
&=&-\oint_{z=\iy}\tau_n(t+[\al]-[z^{-1}])\tau_{n-\ell}(t-[\al]+[z^{-1}])
e^{2\sum_1^{\iy}(\al z)^i/i}\al^2 z^{\ell +1}\frac{dz}{2\pi iz}\\
&=&\al^{1-\ell}\sum_{k=1}^{\iy}\al^k\left(\frac{\pl}{\pl
t_k}-\sum^{\ell-1}_{r=0}(\ell-r)p_r(-\tilde\pl)p_{k-r}(\tilde\pl)\right)
\tau_{n}\circ\tau_{n-\ell},
\end{eqnarray*}
establishing the second relation of (2.8). As for the first one, set
$m=n-1$,
$t\mapsto t-a$ and $t'\mapsto t+a$ in the bilinear identity, and use
(1.4), thus yielding (0.14).

\medbreak

{\bf Step 3:}  To check the formulas (0.12) for $S$, compute
\begin{eqnarray*}
e^{\sum^{\iy}_1t_iz^i}S\chi(z)&=:&\Psi(t,z)\\
&=&e^{\sum^{\iy}_1t_iz^i}\frac{\tau(t-
[z^{-1}])}{\tau(t)}\chi(z)\quad\mbox{(by (0.10))}\\
&=&e^{\sum^{\iy}_1t_iz^i}\sum^{\iy}_{n=0}
\frac{p_n(-\tilde\pl)\tau(t)}{\tau(t)}z^{-n}\chi(z)\\
&=&e^{\sum^{\iy}_1t_iz^i}\sum^{\iy}_0
\frac{p_n(-\tilde\pl)\tau(t)}{\tau(t)}\Lb^{-n}\chi(z).
\end{eqnarray*}
Similarly one checks the formula for $S^{-1}$ using the formulas
for $\Psi^*(t,z)$ in terms of $S^{-1}$ and $\tau(t)$. Finally to
check the formula (0.13) for $L^k$, use the formulas (0.12) for $S$
and $S^{-1}$ (for $\tilde \Lb$, see footnote 1):
\begin{eqnarray*}
L^k&=&S\Lb^kS^{-1}\\
&=&\sum^{\iy}_{i,j\geq
0}\frac{p_i(-\tilde\pl)\tau}{\tau}\Lb
^{-i-j+k}\left(\tilde \Lb\frac{p_j(\tilde\pl)\tau}{\tau}\right)\\
&=&\sum^{\iy}_{i,j\geq
0}\frac{p_i(-\tilde\pl)\tau}{\tau}\left(\tilde \Lb^{-i-j+k+1}
\frac{p_j(\tilde\pl)\tau}{\tau}\right)\Lb^{-i-j+k}\\
&=&\sum_{\ell\geq 0}\left(\sum_{{i,j\geq
0}\atop{i+j=\ell}}\frac{p_i(-\tilde\pl)\tau_np_j(\tilde\pl)\tau_{n+k
-\ell+1})}{\tau_n\tau_{n+k-\ell+1}}\right)_{n\in\BZ}
\Lb^{k-\ell}\\
&=&\sum_{\ell\geq
0}\left(\frac{p_{\ell}(\tilde\pl)
\tau_{n+k-\ell+1}\circ\tau_n}{\tau_{n+k-\ell+1}\tau_n}
\right)_{n\in\BZ}\Lb^{k-\ell}\mbox{\,\,}\quad\mbox{(using (1.13))}
\end{eqnarray*}
yielding (0.13) and (0.15), upon noting,
\begin{eqnarray*}
\mbox{coef}_{\Lb^{k-1}}L^k&=&\left(\frac{p_1(\tilde\pl)\tau_{n+k}
\circ\tau_n}{\tau_{n+k}\tau_n}\right)_{n\in\BZ}=\left(\frac{\pl}{\pl
t_1}\log\frac{\tau_{n+k}}{\tau_n}\right)_{n\in\BZ}\\
\mbox{coef}_{\Lb^0}L^k&=&\left(\frac{p_k(\tilde\pl)\tau_{n+1}
\circ\tau_n}{\tau_{n+1}\tau_n}\right)_{n\in\BZ}=
\left(\frac{\pl}{\pl
t_k}\log\frac{\tau_{n+1}}{\tau_n}\right)_{n\in\BZ}\mbox{\,\,by (2.8)}\\
\mbox{coef}_{\Lb^{-1}}L^k
&=&\left(\frac{p_{k+1}(\tilde\pl)\tau_n
\circ\tau_n}{\tau_n\tau_n}\right)_{n\in\BZ} =\left(\frac{\pl^2}{\pl
t_1\pl t_k}\log\tau_n\right)_{n\in\BZ},\quad\mbox{by (2.7)},
\end{eqnarray*}
concluding the proof of the Theorem 0.1.\qed

\begin{corollary} Setting $\ga(t):=(\tilde \Lb\tau(t)/\tau(t))$, the
wave operator $W(t)$ for the discrete
KP-hierarchy has the following property
$$
(W(t)W^{-1}(t'))_-=0,\quad (W(t)W^{-1}(t'))_0=\frac{\ga(t)}{\ga(t')}.
$$
\end{corollary}

\proof That $h(t,t')=W(t)W^{-1}(t')\in\DR_+$ was shown in Lemma 2.2.
Concerning its diagonal $h_0$, we deduce from (2.3)
that\footnote{$M_0:=$~diagonal part of $M$.}
$$
\frac{\pl}{\pl t_k}\log h_0=(L^k(t))_0,\quad
\frac{\pl}{\pl t_k}\log h_0=-(L^k(t'))_0,\quad\mbox{with
$h_0(t,t)=I.$}
$$
Note that $\ga(t)/\ga(t')$ satisfies the same differential equations
as $h_0(t)$ with the same initial condition, upon using (0.15):
\begin{eqnarray*}
\left(\frac{\pl}{\pl t_k}\log\frac{\ga(t)}{\ga(t')}\right)_n&=&
\frac{\pl}{\pl t_k}\log\frac{\tau_{n+1}(t)}{\tau_n(t)}=L^k(t)_{nn}\\
\left(\frac{\pl}{\pl t'_k}\log\frac{\ga(t)}{\ga(t')}\right)_n&=&
-\frac{\pl}{\pl
t'_k}\log\frac{\tau_{n+1}(t')}{\tau_n(t')}=-L^k(t')_{nn},
\end{eqnarray*}
with $\ga(t)/\ga(t')\Biggl|_{t=t'}=I$.

\section{Sequences of $\tau$-functions, flags and the discrete
KP equation}

In this section, we prove Theorem 0.2; it will be broken up into three
propositions: the first one is very similar to the analogous statement
for the KP theory (see \cite{DJKM,vM}). One could make an argument
unifying both cases, in the context of Lie theory. The second statement
uses Grassmannian technology.

\begin{proposition} The following equivalences (i)
$\Longleftrightarrow$ (ii) $\Longleftrightarrow$ (iii) hold.
\end{proposition}

\proof (i) $\Rg$ (ii) was already shown in Theorem 0.1. Regarding
the converse (ii) $\Rg$ (i), we show vectors $\Psi(t,z)$ and
$\Psi^*(t,z)$ having the asymptotics (0.8) and satisfying the bilinear
identity (ii) are discrete
KP-hierarchy vectors.

\medbreak

The point of the proof is to show that the matrices $S$ and $T^t\in
I+\DR_-$ defined through
$$
\Psi(t,z)=:e^{\sum_1^{\iy}t_iz^i}S\chi(z),\quad
\Psi^*(t,z)=:e^{-\sum_1^{\iy}t_iz^i}T\chi^*(z)
$$
satisfy the vector fields (0.6) with $T^t=S^{-1}$.

\medbreak

\noindent{\bf Step 1.} $T^t=S^{-1}$.

Assuming the bilinear identities (assumption (ii) of Theorem 0.2),
\begin{eqnarray*}
0&=&\left(\oint_{z=\iy}\Psi(t,z)\otimes\Psi^*(t,z)\frac{dz}{2\pi
iz}\right)_-\\
&=&\left(\oint_{z=\iy}e^{\sum_1^{\iy}t_iz^i}S  \,\chi(z)\otimes
e^{-\sum_1^{\iy}t_iz^i}T\,\chi(z^{-1})\frac{dz}{2\pi iz}
\right)_-    \\
&=&(ST^{\top})_-,\quad\mbox{by (2.1)}
\end{eqnarray*}
but since $S,T^t\in I+\DR_-$, $ST^t=I$, yielding $T^t=S^{-1}$.

\medbreak

\noindent{\bf Step 2.} $W(t)W^{-1}(t')\in\DR_+$, upon defining
$W(t):=S(t)e^{\sum t_i\Lb^i}$.

According to the bilinear identity, the left hand side of

$$\oint_{z=\iy}\Psi(t,z)\otimes\Psi^*(t',z)\frac{dz}{2\pi
iz}\hspace{5cm}$$
\begin{eqnarray*}
&=&\oint_{z=\iy}e^{\sum t_iz^i}S \, \chi(z)\otimes
e^{-\sum_1^{\iy}t'_iz^i}(S^{-1})
^{\top}\chi(z^{-1})\frac{dz}{2\pi
iz}\\
&=&\oint_{z=\iy}S(t)e^{\sum t_i\Lb^i}\,
\chi(z)\otimes (S^{-1}(t'))^{\top}e^{-\sum
t'_i\Lb^{\top -i}} \chi(z^{-1})\frac{dz}{2\pi iz}\\ &=&S(t)e^{\sum
t_i\Lb^i}e^{-\sum t'_i
\Lb^i}S^{-1}(t'),\quad\mbox{using Lemma 2.1}\\ &=&W(t)W^{-1}(t');
\end{eqnarray*}
 belongs to $\DR_+$, and hence so is the right hand side.

\medbreak

\noindent{\bf Step 3.}
\begin{eqnarray*}
\left(\frac{\pl}{\pl t_n}-(L^n)_+\right)\Psi(t,z)
&=&\left(\frac{\pl}{\pl
t_n}-(L^n)_+\right)S\chi(z)e^{\sum_1^{\iy}t_iz^i}\\
&=&\left(\frac{\pl S}{\pl t_n}-(L^n)_+S+S\,z^n\right)\chi(z)
e^{\sum_1^{\iy}t_iz^i}\\
&=&\left(\frac{\pl S}{\pl
t_n}-(L^n)_+S+S\,\Lb^n(S^{-1}S)\right)
\chi(z)e^{\sum_1^{\iy}t_iz^i}\\
&=&\left(\frac{\pl S}{\pl
t_n}-(L^n)_+S+L^nS\right)\chi(z)
e^{\sum_1^{\iy}t_iz^i}\\
&=&\left(\frac{\pl S}{\pl
t_n}+(L^n)_-S)\right)\chi(z)e^{\sum_1^{\iy}t_iz^i}.
\end{eqnarray*}

\medbreak

\noindent{\bf Step 4.} From $W(t)W^{-1}(t')\in\DR_+$, since
$\DR_+$ is an algebra, deduce
\begin{eqnarray*}
\DR_+&\ni&\left(\left(\frac{\pl}{\pl t_n}-(L^n)_+\right)W(t)\right)
W^{-1}(t')\Biggl|_{t'=t}\\
&=&\oint_{z=\iy}\left(\frac{\pl}{\pl t_n}-(L^n)_+\right)\Psi(t,z)
\otimes\Psi^*(t,z)\frac{dz}{2\pi iz},\quad\mbox{by step 2}\\
&=&\oint_{z=\iy}\left(\frac{\pl S(t)}{\pl t_n}+(L^n)_-S(t)\right)
\chi(z)e^{\sum_1^{\iy}t_iz^i}\otimes(S^{\top}(t))^{-1}
\chi(z^{-1})e^{-\sum_1^{\iy}t_iz^i}\frac{dz}{2\pi iz},\\
& &\hspace{7cm}\quad\mbox{by step 3}\\
&=&\left(\frac{\pl S(t)}{\pl t_n}+(L^n)_-S(t)\right)S(t)^{-1},\quad
\mbox{by Lemma 2.1}
\end{eqnarray*}
and thus, since $S\in I+\DR_-$ and $\DR_-$ is an algebra,
$$
\left(\frac{\pl S(t)}{\pl
t_n}+(L^n)_-S(t)\right)S(t)^{-1}\in\DR_+\cap\DR_-=0;
$$
therefore, we have the discrete
KP-hierarchy equations on $S$
$$
\frac{\pl S(t)}{\pl t_n}+(L^n)_-S=0,\quad n=1,2,...,
$$
and on $L=S\Lb S^{-1}$,
$$
\frac{\pl L}{\pl t_n}=[-(L^n)_-,L],
$$
ending the proof that (ii) $\Rg$ (i).\qed

\medbreak

Finally (ii) $\Longleftrightarrow$ (iii) upon using the
equivalence (i) $\Longleftrightarrow$ (ii) and the
$\tau$-function representation (0.10) of $\Psi$ and $\Psi^*$, shown in
Theorem 0.1; this establishes Proposition 3.1.

With each component of the wave vector $\Psi$, given in (0.10), or,
what is the same, with each component of the $\tau$-vector, we
associate a sequence of infinite-dimensional planes in the
Grassmannian $Gr^{(n)}$
\bea
\WR_n&=&\mbox{ span}_{\BC}\left\{\left(\frac{\pl}{\pl t_1}
\right)^k\Psi_n(t,z),~~k=0,1,2,...\right\}\nonumber \\
&=&e^{\sum_1^{\iy}t_i z^i}\mbox{ span}_{\BC}\left\{\left(
\frac{\pl}{\pl
t_1}+z\right)^k\psi_n(t,z),~~k=0,1,2,...\right\}\nonumber\\
&=:& e^{\sum_1^{\iy}t_i z^i} \WR_n^t.
\eea
and planes
\be
\WR^\ast_n=\mbox{ span}_{\BC}\left\{\frac{1}{z}\left(\frac{\pl}{\pl t_1}
\right)^k \Psi^{\ast}_{n-1}(t,z),~~k=0,1,2,...\right\},
\ee which are the orthogonal complements of $\WR_n$ in $Gr^{(n)}$, by
the residue pairing
\be
\la f,g\ra_{\iy} :=\oint_{z=\iy} f(z) g(z) \frac{dz}{2\pi i}.
\ee

\begin{proposition} (ii) $\Longleftrightarrow$ (iv)
$\Longleftrightarrow$ (v) holds.
\end{proposition}

\proof The inclusion $...\supset \WR_{n-1}\supset \WR_n\supset
\WR_{n+1}\supset ...$ in (iv) implies that $\WR_n$,
given by (3.1) and (0.10), is also given by
$$
\WR_n=\mbox{span}_{\BC}\{\Psi_n(t,z),\Psi_{n+1}(t,z),...\}.
$$
Moreover the inclusions $...\supset \WR_n\supset
\WR_{n+1}\supset ...$ imply, by orthogonality, the inclusions $...
\subset \WR_n^*\subset \WR^*_{n+1}\subset ...$, and thus $\WR^*_n$,
given by (3.2) and (0.10) and thus specified by $\Psi^*_{n-1}$ and
$\tau_n$, is also given by
$$
\WR^*_n=\{\frac{\Psi^*_{n-1}(t,z)}{z},\frac{\Psi^*_{n-2}(t,z)}{z},...\}.
$$
The formula (0.10) for $\Psi_n$ and $\Psi^*_{n-1}$
 imply the
bilinear identities (1.1), since each $\tau_n$ is a
$\tau$-function, yielding $\WR_n^{\ast}
=\WR_n^{\bot}$, with respect to the residue pairing and so:
$$
\la \Psi_n(t,z),\frac{\Psi^*_{n-1}(t',z)}{z}\ra_{\iy}
=\oint_{z=\iy}\Psi_n(t,z)\Psi^*_{n-1}(t',z)\frac{dz}{2\pi iz}=0.
$$
Since
$$
\WR_n\subset \WR_{m+1}=(\WR^*_{m+1})^*,\quad\mbox{all $n>m$}
$$
we have the orthogonality $\WR_n\perp \WR^*_{m+1}$ for all $n>m$, with
respect to the residue pairing; since $\Psi_n(t,z) \in \WR_n,
~\frac{\Psi^*_m(t',z)}{z}\in \WR^*_{m+1}(t',z)$, we have
\be
0=\la \Psi_n(t,z),\frac{\Psi^*_m(t',z)}{z}\ra_{\iy}
=\oint_{z=\iy}\Psi_n(t,z)\Psi^*_m(t',z)\frac{dz}{2\pi
iz},\quad\mbox{all $n>m$,}
\ee
which is (ii).

Now assume (ii); then, for fixed $n>m$, we have
$$
0=\oint_{z=\iy}\left(\frac{\pl}{\pl t_1}\right)^k
\Psi_n(t,z)\left(\frac{\pl}{\pl
t'_1}\right)^{\ell}\Psi^*_m(t',z)\frac{dz}{2\pi iz},\quad n>m
$$
and thus by (3.1) and (3.2),
$$
\WR_n\subseteq (\WR^{\ast}_{m+1})^{\ast}=\WR_{m+1},\quad\mbox{for
$n>m$,}
$$
which implies the flag condition $...\supset \WR_{n-1}\supset
\WR_n\supset \WR_{n+1}\supset ...$, stated in (iv).

(iv) $\Longleftrightarrow$ (v), follows from the equivalence of
(i) and (iii) in Proposition 1.1, by setting $\tau_1:=\tau_{n-1}$,
$\tau_2=\tau_n$,
$\tilde \WR_1=z^{-n+1}\WR_{n-1}$ and $\tilde \WR_2=z^{-n}\WR_n$ and
noting
$$
z(z^{-n}\WR_n)\subset(z^{-n+1}\WR_{n-1}),\quad\mbox{i.e. $\WR_n\subset
\WR_{n-1}$},
$$
concluding the proof of the proposition.\qed

\begin{proposition} (v)
$\Longleftrightarrow$ (vi) holds.
\end{proposition}

\proof

{\bf\noindent Step 1.} For a given $n\in\BZ$, statement (v), namely
$$
R_k^{(n)}:=\{p_{k-1}(-\tilde\pl)\tau_n,\tau_{n+1}\}+\tau_{n+1}p_k(-\tilde
\pl)\tau_n-\tau_np_k(-\tilde\pl)\tau_{n+1}=0,\quad k\geq 2
$$
implies
$$R_k^{(n)'}=\left(\frac{\pl}{\pl
t_k}-p_k(\tilde\pl)\right)\tau_{n+1}\circ\tau_n=0,\quad k\geq 2.
$$
Since $R_k^{(n)}$ are the Taylor coefficients of relation (v) in
Theorem 0.2, statement (v)$_n$ is equivalent to (iv)$_n$ (i.e.
$\WR_n\supset \WR_{n+1}$). The latter is equivalent to the bilinear
identity (iii)$_n$ (i.e., (0.18) with $n\rg n+1$ and $m\rg n-1$).
According to the arguments used in the proof of Theorem 0.1,
(iii)$_n$ implies $R_k^{(n)'}=0$.

\medbreak

{\bf\noindent Step 2.} The converse holds, because, upon using an
inductive argument,
$$
R_k^{(n)}=\al R_k^{(n)'}+\mbox{\,partials of
$(R_1^{(n)'},...,R_{k-1}^{(n)'})$};
$$
thus the vanishing of the $R_1^{(n)'},...,R_k^{(n)'}$ implies the
vanishing of $R_k^{(n)}$.\qed

\bigbreak

\begin{theorem}
Every 1-Toda lattice is equivalent to a 2-Toda lattice.
\end{theorem}

\proof The 1-Toda theory implies for $S_1:=S\in I+{\cal
D}_-$,
$L_1:=L$
$$
\frac{\pl S_1}{\pl t_n}=-(L_1^n)_-S_1(t),\quad\mbox{where
$L_1=S_1\Lb S_1^{-1}.$}
$$
Then, in view of the 2-Toda theory, define $S_2(t)\in{\cal D}_+$ by
means of the differential equations
$$
\frac{\pl S_2(t)}{\pl t_n}=(L_1^n)_+S_2(t),\quad n=1,2,...,
$$
with initial condition $S_2(0)=$ (an invertible element
$d_+\in{\cal D}_+$). Then define\footnote{The first index in
$L_{1,2}$ and $S_{1,2}$ corresponds to the upper-sign.}
$S_{1,2}(t,s)$ and
$L_{1,2}=S_{1,2}\Lb^{\pm 1}S_{1,2}^{-1}$, flowing according to
the commuting differential equations
\be
\frac{\pl
S_{1,2}(t,s)}{\pl
s_n}=\pm(L_2^n(t,s))_{\mp}S_{1,2}(t,s)\quad\mbox{with}\quad
S_{1,2}(t,0)=S_{1,2}(t).
\ee
$S_{1,2}(t,s)$ satisfies the $t$-equations of 2-Toda for $s=0$, by
construction; now we must check that this holds for $s\neq 0$;
therefore, set
\be
F_{1,2}^{(n)}(t,s)=\frac{\pl
S_{1,2}}{\pl t_n}(t,s)\pm(L^n_1(t,s))_{\mp}S_{1,2}(t,s),
\quad\mbox{for
$n=1,2,...$}
\ee
Compute, using (3.5) and $[\pl/\pl t_n,\pl/\pl s_n]=0$, the
system of two differential equations
$$
\frac{\pl
F_{1,2}^{(n)}}{\pl
s_k}(t,s)=\pm[F_{2,1}^{(n)}S_2^{-1},L_2^k]_{\mp}S_{1,2}\pm(L_2^k)_{\mp}
F_{1,2}^{(n)},\quad k,n=1,2,...;
$$
since $F_{1,2}^{(n)}(t,0)=0$, we have $F_{1,2}^{(n)}(t,s)=0$ for all
$s$. Thus, by (3.5) and (3.6), $S_{1,2}(t,s)$ flow according to
2-Toda.\qed

\medbreak

\section{Discrete KP-solutions generated by vertex operators}

An important construction leading to Toda solutions is contained
in Theorem 0.3, which is based on the following Lemma:

\begin{lemma}
 Particular solutions to equation
\be\{ \tau_1
(t-[z^{-1}]),\tau_2 (t)\} + z (\tau_1 (t-[z^{-1}]) \tau_2 (t)
- \tau_2 (t-[z^{-1}])\tau_1 (t)) = 0 \ee
are given, for arbitrary $\lambda \in
\BC^\ast$, by pairs $(\tau_1,\tau_2)$, defined by:
\be
\tau_2 (t) = \left(\int X (t,\lambda)\nu(\lb)d\lb\right) \tau_1 (t) =
\int e^{\sum t_i
\lambda^i}
\tau_1 (t-[\lambda^{-1}])\nu(\lb)d\lb,
\ee
or
\be
\tau_1 (t) =\left(\int X (-t,\lambda)\nu'(\lb)d\lb\right) \tau_2 (t) =
\int e^{-\sum t_i \lambda^i}
\tau_2 (t+[\lambda^{-1}])\nu'(\lb)d\lb.
\ee
\end{lemma}

\proof  Using
$$e^{-\sum^\iy_1 {1 \over i}({\lambda \over z})^i } = 1 -
{\lambda \over z},$$ it suffices to check,before even integrating,
that $\tau_2 (t)=X(t,\lb)\tau_1(t)$ satisfies the above equation
(4.1)
\bea
&&e^{-\sum t_i \lambda^i} \left(\{\tau_1
(t-[z^{-1}]),\tau_2(t)\}
 + z (\tau_1 (t-[z^{-1}])\tau_2(t) - \tau_2(t-[z^{-1}]) \tau_1
(t))\right) \nonumber\\
&&~~~~= e^{-\sum t_i \lambda^i}\{\tau_1 (t-[z^{-1}]),
e^{\sum t_i \lambda^i} \tau_1 (t-[\lambda^{-1}])\} \nonumber
\\  &&~~~~~~~~~+ z
(\tau_1 (t-[z^{-1}]) \tau_1 (t-[\lambda^{-1}]) - (1-{\lambda
\over z}) \tau_1 (t) \tau_1  (t-[z^{-1}] -
\lambda^{-1}]))\nonumber  \\ &&~~~~= \{\tau_1
(t-[z^{-1}]),\tau_1 (t-[\lambda^{-1}])\}\nonumber
\\ &&~~~~~~~~~+
(z-\lambda) (\tau_1 (t-[z^{-1}])\tau_1 (t-[\lambda^{-1}]) -
\tau_1 (t) \tau_1 (t-[z^{-1}]-[\lambda^{-1}]))\nonumber    \\
&&~~~~= 0,\nonumber
\eea
using the differential Fay identity (1.3) for the  $\tau$-function
$\tau_1$; a similar proof works for the second solution, given by
$\tau_1 (t)=X(-t,\lb)\tau_2(t)$. Since equation (4.1) is linear in
$\tau_1(t)$, and also in $\tau_2(t)$, the equation remains valid after
integrating with regard to $\lb$.
\qed

\noindent\underline{\em Proof of Theorem 0.3}: Note, from the
definition of $\tau_{\pm n}$ in Theorem 3, that each $\tau_n$ is
defined inductively by
$$
\tau_{n+1} =\int  X(t,\lambda) \nu_n(\lb) d\lb ~\tau_n\mbox{ and
}\tau_{-n-1}= \int X(-t,\lambda) \nu_{-n-1}(\lb) d\lb ~\tau_{-n};
$$
thus by Lemma 4.1, the functions $\tau_{n+1}$ and
$\tau_n$ are a solution of equation (v) of Theorem 0.2. Therefore,
theorem 0.2 implies that the
$\tau_n$'s form a $\tau$-vector of the discrete KP hierarchy.\qed

\section{Example of vertex generated solutions: the $q$-KP equation}

Consider the class of $q$-pseudo-difference operators,
with $y$-dependent coefficients, acting on functions $f(y)$
$$
\DR_q=\{ \sum a_i(y) D^i\},~~\mbox{with}~~Df(y):=f(qy).
$$
and the $q$-derivative $D_q$, defined by
$$
D_qf(y):=\frac{f(qy)-f(y)}{(q-1)y}=-\lb(y)(D-1)f(y),
~\mbox{with}~\lb(y):=-\frac{1}{(q-1)y};
$$
Consider the following $q$-pseudo-difference operators
$$
Q=D+u_0(x)D^0+u_{-1}D^{-1}+...\mbox{
and }~~Q_q=D_q+v_0(x)D_q^0+v_{-1}(x)D_q^{-1}+...
$$
and the following $q$-deformations, which were proposed respectively by
E. Frenkel \cite{F} and Khesin, Lyubashenko and Roger \cite{KLR}, for
$n=1,2,...$:
\be
{\pl Q\over\pl t_n}=\bigl[\left(Q^n\right)_+,Q\bigr]
~~~~~~~~~~~~~~~ ~~~~~~~\mbox{{\em (Frenkel system)}}
\ee
\be
{\pl Q_q\over\pl
t_n}=\bigl[\left(Q_q^n\right)_+,Q_q\bigr],
~~~~~~~~~~~~~~~~~~~ \mbox{{\em (KLR system)}}
\ee
where $(~)_+$ and $(~)_-$ refer to the $q$-difference and
strictly $q$-pseudo-differential part of $(~)$. 
Define
\be
 c(x) =
\left({(1-q) x \over 1-q} , {(1-q)^2 x^2 \over 2(1-q^2)} ,
{(1-q)^3 x^3 \over 3 (1-q^3)}, ... \right)\in \BC^{\iy}\ \ \ \mbox{and}
\
\
\
\lambda^{-1}_n = (1-q) x q^{n-1} ,
\ee
one checks for $n \geq 1$, $D^n \lb_0(x)=\lb_n(x)$, and
\bea
D^n c(x)&=&c(x)-\sum_1^n[\lb_i^{-1}(x)]\nonumber\\
D^{-n} c(x)&=&c(x)+\sum_1^{n}[\lb_{-i+1}^{-1}(x)]
\eea
Details about these theorems were reported in a joint work with E.
Horozov\cite{AHV}.

\begin{theorem}
There is an algebra isomorphism
$$
\hat {}~
: \DR_q  \lrg    \DR ,
$$
which maps the Frenkel and KLR system into the discrete
KP-hierarchy
\be
{\pl L\over\pl t_n}=\bigl[\left(L^n\right)_+,L \bigr],\quad
n=1,2,.. .
\ee
\end{theorem}
\bigbreak

\begin{theorem} The matrices
$$L=\Lb+\sum_{-\iy<\ell\leq 0}\diag\left(
\frac{p_{1-\ell}(\tilde\pl)\tau_{n+\ell+1}\circ\tau_n}{\tau_{
n+\ell+1}\tau_n}\right)_{n\in\BZ}\Lb^{\ell}
$$
and
$$
\tilde L=\vr L\vr^{-1}
$$
with $\vr$ as in (5.11), $\tau_0=\tau(c(x)+t)$ and
\bea
\tau_n&=&X(t,\lb_n)...X(t,\lb_1)\tau(c(x)+t)\nonumber\\
&=&r_n(\lb)\left(
\prod^n_{k=1}e^{\sum^{\iy}_{i=1}t_i\lb^i_k}\right)D^n\tau(c(x)+t)
\eea
\bea
\tau_{-n}&=&X(-t,\lb_{-n+1})...X(-t,\lb_0)\tau(c(x)+t)\nonumber\\
&=&r_{-n}(\lb)\left(
\prod^n_{k=1}e^{-\sum^{\iy}_{i=1}t_i\lb^i_
{-k+1}}\right)D^{-n}\tau(c(x)+t) \nonumber
\eea
transform, using the map $\hat {}$, respectively into solutions to
the $q$-KP deformations (5.1) and (5.2) of
$$
Q=D+\sum_{-\iy<i\leq 0}a_i(y)D^i\quad\mbox{or}\quad
Q_q=D_q+\sum_{-\iy<i\leq 0}b_i(y)D^i_q,
$$
where the $b_i$ are related to the $a_i$ by (5.12)
and\footnote{$\pi(k)=$ parity of $k=1$, when $k$ is even, and
$=-1$, when $k$ is odd.}
$$
a_{\ell}(y)=\mbox{\,polynomial in}\left\{
\begin{tabular}{l}
$\displaystyle{\frac{\pl^k}{\pl
t_{i_1}...\pl
t_{i_k}}\log\left(\tau(c(y)+t)^{\pi(k)}D^{\ell+1}\tau(c(y)+t)\right)}$
for
$k\geq 2$\\
 \\
$\displaystyle{\sum^{\ell+1}_{i=1}\lb_i^j(y)+\frac{\pl}{\pl
t_j}\log\frac{D^{\ell+1}\tau(c(y)+t)}{\tau(c(y)+t)}}$, for $k=1$
\end{tabular}\right.
$$
\end{theorem}

The proofs of these theorems, which rely heavily on the next lemma,
will be given later. In an elegant recent paper, Iliev \cite{I} has
obtained $q$-bilinear identities and $q$-tau functions, as well,
purely within the KP theory.

Consider an appropriate space of functions $f(y)$ representable by
``Fourier" series
$$
f(y)=\sum_{-\iy}^{\iy}f_n\varphi_n(y)
$$
in the  basis\footnote{The $\delta$-function $\dt(z):=\sum_{i \in \BZ}z^i$;
enjoys the property $f(za)\dt(z)=f(a)\dt(z)$}
$\varphi_n(y):=\dt(q^{-n}x^{-1}y)$ for fixed $q\neq 1$, and a
parameter $x \in \BR$. Also, remember
\be
\lb_i:=D^i\lb_0=\lb(xq^i).
\ee

\begin{lemma} Then the Fourier transform,
$$
f\longmapsto \FR f=(...,f_n,...)_{n\in \BZ},
$$
induces an algebra isomorphism $\hat {}$, mapping $D$-operators
into $\Lambda$-operators
\bea
\hat {}~ : \DR_q &\lrg  & \DR   \nonumber\\
\sum_i a_i(y)D^i  &\longmapsto &
\sum \hat a_i  \Lb^i:=\sum_i\mbox{ diag}\left(
...,a_i(xq^n),...\right)_{n
\in \BZ}\Lb^i.
\eea
Moreover
\bea
\sum_{i=0}^n b_i(y) D_q^i=\sum_{i=0}^n a_i(y)(-\lb D)^i &\longmapsto &
\vr  \left( \sum_{i=0}^n \hat a_i  \Lb^i \right)  \vr^{-1}  ,
\eea
where the $\Lb$-operator in brackets is monic,
with\footnote{with $[j]:=\frac{1-q^j}{1-q}$ and $\bigl[ \stackrel{n}{k}
\bigr]:=\frac{[n]~[n-1]~...[n-k+1]}{[k]~[k-1]~...[1]}$}
\be
\hat\lb=(...,\lb_{-1}(x),\lb_0(x),\lb_1(x),...)=(...,D^{-1}\lb,\lb,D\lb,...)
\ee
\be
\vr:=\mbox{diag}~\left(...,\lb_{-2}\lb_{-1},-\lb_{-1},1,
-\frac{1}{\lb_0},\frac{1}{\lb_0\lb_1},
-\frac{1}{\lb_0 \lb_1 \lb_2},...\right)~~\mbox{with}~ \vr_0=1,
\ee
\be
a_i(y):=\sum_{0\leq k \leq
n-i}\frac{\bigl[\stackrel{k+i}{k}\bigr]}{(-y(q-1)q^i)^k} b_{k+i}(y).
\ee
\end{lemma}

\proof The operators $D$ and multiplication by a function
$a(y)$ act on basis elements, as follows:
$$
D\varphi_n(y)=\varphi_{n-1}(y)~~\mbox{ and }~~a(y)
\varphi_n(y)=a(xq^n)\varphi_n(y).
$$
Therefore $D^k$ and $a(y)$ act on functions $f(y)$, as
\bea
f(y)=\sum_{n \in \BZ} f_n\varphi_n(y) \longmapsto D^kf(y)&=&\sum_{n
\in \BZ}f_n D^k \varphi_n(y)\nonumber\\
&=& \sum_{n \in \BZ}f_n \varphi_{n-k}(y) \nonumber\\
&=& \sum_{n \in \BZ}f_{n+k} \varphi_{n}(y),
\eea
and
\bea
f(y)=\sum_{n \in \BZ} f_n\varphi_n(y) \longmapsto a(y)f(y)&=&\sum_{n
\in \BZ}f_n a(y) \varphi_n(y) \nonumber\\
&=&\sum_{n
\in \BZ}f_n a(xq^n) \varphi_n(y),
\eea
from which it follows that
\be
 (D^k)\hat{}    =\Lb^k
\ee
\be
\hat a(y)=\mbox{diag}~(...,a(xq^n),...)_{n \in \BZ}.
\ee
To establish the algebra isomorphism (5.8), one checks that

\bea
\left( a(y) D^i \right)\hat{} ~\left( b(y) D^j \right)\hat{} &=&\hat
a(y)
\Lb^i ~ \hat b(y) \Lb^j  \nonumber\\ &=&\hat a(y) \left( \Lb^i \hat
b(y) \Lb^{-i} \right) \Lb^{i+j}
\nonumber\\
&=&\mbox{diag}(...,a(xq^n)b(xq^{n+i}),...)_{n \in \BZ}\Lb^{i+j}
\nonumber\\
&=& \left( a(y)b(yq^i) D^{i+j}   \right)\hat{}\nonumber\\
&=&\left( a(y) D^i ~~b(y) D^j \right)\hat{}.
\eea
Using the inductively established identity
$$
D_q^n=\frac{1}{y^n (q-1)^n q^{\frac{n(n-1)}{2}}}\sum_{k=0}^n(-1)^k
q^{k(k-1)/2}\Bigl[\stackrel{n}{k}\Bigr] D^{n-k},
$$
the first identity (5.9) is immediate.

Then, using, by virtue of (5.10) and (5.11), $ \hat \lb \Lb=-\vr \Lb
\vr^{-1}$ and
$\vr
\hat a
\vr^{-1}=\hat a$ (since $\hat a$ is diagonal), one computes, using
 the established isomorphism,
\bea
 \left(  a_i(y)( -\lb(y) D)^i\right)\hat{}&=&  \hat a_i \left(
-\hat\lb
\hat D \right)^i
\nonumber\\
&=&\hat a_i \left( -\hat \lb \Lb \right)^i\nonumber\\
&=& \hat a_i \left(\vr \Lb \vr^{-1} \right)^i \nonumber\\
&=&\vr  \left(\hat a_i \Lb^i  \right)   \vr^{-1}
\eea
establishing (5.9).   \qed

\noindent\underline{\em Proof of Theorem 5.1}:
Indeed the Frenkel system (5.1) maps at once into (5.5), whereas,
using (5.9), the KLR-system maps into
\bea
{\pl \vr L \vr^{-1} \over \pl
t_n}&=&\bigl[\left(\vr L^n \vr^{-1}\right)_+,\vr L \vr^{-1} \bigr] \\
&=& \vr   \bigl[\left(L^n\right)_+,L\bigr]   \vr^{-1},
\eea
which upon conjugation by $\vr$ leads to (5.5) as well.\qed

\bigbreak

\underline{\sl Proof of Theorem 5.2}: From Theorem 0.3, it follows that
$L$ with the
$\tau_n$'s defined by (5.6), satisfies the Toda lattice; the second
equality in (5.6) follows from (5.4). According to Lemma 1.3,
$$
\frac{p_{1-\ell}(\tilde\pl)\tau_{n+\ell+1}\circ\tau_n}{\tau_{n+\ell+1}
\tau_n}=\mbox{\,a
polynomial in
$\left(\frac{\pl^k}{\pl
t_{i_1}...\pl t_{i_k}}\log(\tau_{n+\ell+1}\tau_n^{\pi(k)})\right)$},
$$
where by (5.6), for $k\geq 2$,

\medbreak\noindent
$\displaystyle{\left(\frac{\pl^k}{\pl
t_{i_1}...\pl t_{i_k}}\log(\tau_{n+\ell+1}\tau_n
^{\pi (k)})\right)_{n\in\BZ}}$
\begin{eqnarray*}
&=&\left(D^n\frac{\pl^k}{\pl t_{i_1}...\pl
t_{i_k}}\log\left(\tau(c(y)+t)^{\pi(k)}D^{\ell+1}\tau(c(y)+t)\right)
\right)_{n\in\BZ}\\
&=&\left(\frac{\pl^k}{\pl
t_{i_1}...\pl
t_{i_k}}\log\tau(c(y)+t)^{\pi(k)}D^{\ell+1}
\tau(c(y)+t)\right)^{\wedge},
\end{eqnarray*}
and

$\displaystyle{\left(\frac{\pl}{\pl
t_j}\log\frac{\tau_{n+\ell+1}}{\tau_n}\right)_{n\in\BZ}}$
\begin{eqnarray*}
&=&\left(\frac{\pl}{\pl
t_j}\log\frac{\left(
\displaystyle{\prod^{n+\ell+1}_{\al=1}e^{\sum^{\iy}_{i=1}}t_i\lb^i_{\al}}
\right)D^{n+\ell+1}\tau(c(y)+t)}{\left(\displaystyle{
\prod^n_{\al=1}e^{\sum^{\iy}_{i=1}}t_i\lb^i_{\al}}
\right)D^n\tau(c(y)+t)}\right)_{n\in\BZ}\\
&=&\left(\sum_{\al=n+1}^{n+\ell+1}\lb^j_{\al}(y)+
\frac{\pl}{\pl t_j}
\log\frac{D^{n+\ell+1}\tau(c(y)+t)}{D^n\tau(c(y)+t)}
\right)_{n\in\BZ}\\
&=&\left(D^n\left(\sum_{i=1}^{\ell+1}\lb^j_i(y)+
\frac{\pl}{\pl t_j}
\log\frac{D^{\ell+1}\tau(c(y)+t)}{\tau(c(y)+t)}
\right)\right)_{n\in\BZ}\\
&=&\left(\sum_{i=1}^{\ell+1}\lb^j_i(y)+
\frac{\pl}{\pl t_j}
\log\frac{D^{\ell+1}\tau(c(y)+t)}{\tau(c(y)+t)}
\right)^{\wedge},
\end{eqnarray*}
establishing Theorem 5.2.\qed

\noindent\underline{\em Remark}: Note the $\vr$-conjugation has no
counterpart in $\DR_q$-world.

Defining the simple $q$-vertex operators:
$$
X_q(x,t,z):=e_q^{xz}X(t,z)\quad\mbox{and}\quad\tilde X_q
(x,t,z):=(e_q^{xz})^{-1}X(-t,z)
$$
in terms of the vertex operator (6.1) and the $q$-exponential
$\displaystyle{e^x_q=e^{\sum_1^{\iy}\frac{(1-q)^kx^k}{k(1-q^k)}}}$ we
now state:

\begin{corollary}Any K-P $\tau$-function leads to a
$q$-K-P $\tau$-function $\tau(c(x)+t)$ satisfying
$q$-bilinear relations below for all $x\in\BR$,
$t,t'\in\BC^{\iy}$ and all $n>m$, which tends to the
standard K-P bilinear identity when $q$ goes to 1:
$$
\oint_{z=\iy}D^n(X_q(x,t,z)\tau(c(x)+t))D^{m+1}(\tilde
X_q(x,t',z)\tau(c(x)+t')dz=0
$$
$$
\lrg\int_{z=\iy}X(t,z)\tau(\bar x+t)X(t',z)\tau(\bar x+t')dz=0.
$$
\end{corollary}

\proof The $\tau$-functions $\tau_n$ defined in Theorem 5.2 satisfy
the usual bilinear identity (0.18), and so, using the following
identity
\begin{eqnarray*}
\frac{z^{n-m-1}}{\prod^n_{k=m+2}(-\lb)^k}\prod^n_{k=m+2}
e^{-\sum^{\iy}_{i=1}\frac{1}{i}\left(\frac{\lb_k}{z}\right)^i}&=&
\prod^n_{k=m+2}\left(1-\frac{z}{\lb_k}\right)\\
&=&\prod^n_{k=m+2}e^{-\sum^{\iy}_{i=1}\frac{1}{i}\left(\frac{z}{\lb_k}
\right)^i}\\
&=&D^ne_q^{xz}D^{m+1}(e_q^{xz})^{-1}
\end{eqnarray*}
in computing $\tau_n(t-[z^{-1}])$ in the usual bilinear identity
yields, up to a multiplicative factor $\al(\lb,\nu)$:
\begin{eqnarray*}
& &\al(\lb,\nu)\oint_{z=\iy}\tau_n(t-[z^{-1}])\tau_{m+1}(t'+
[z^{-1}])e^{\sum_1^{\iy}(t_i-t'_i)z^i}z^{n-m}\frac{dz}{z}\\
&=&\oint_{z=\iy}\tau(c(x)+t-[z^{-1}]-\sum^n_1[\lb_i^{-1}])\tau(c(x)+t'
+[z^{-1}]+\sum_1^{m+1}[\lb_i^{-1}])\\
& &\quad\quad\quad\quad\prod^n_{k=m+2}\left(1-\frac{z}{\lb_k}
\right)e^{\sum^{\iy}_1(t_i-t'_i)z^i}dz\\
&=&\oint_{z=\iy}D^n(X_q(x,t,z)\tau(c(x)+t))D^{m+1}(\tilde
X_q(x,t',z)\tau(c(x)+t'))dz=0,
\end{eqnarray*}
the latter tending as $q\rg 1$ to the usual KP bilinear identity,
upon using (5.3).\qed

\begin{corollary}If we take $\tau_0(t)=\tau(c(x)+t)$
in Theorem 5.2, with $\tau(t)$ a $N$-KdV $\tau$-function, i.e.,
$\displaystyle{\pl\tau  /  \pl t_{iN}=0}$,
$i=1,2,...$, then
\be
(L^N)=(L^N)_+\quad\mbox{and}\quad\tilde L^N=(\tilde L^N)_+
\ee
yielding the $N$-Frenkel and $N$-KLR system:
\be
Q^N=(Q^N)_+\quad\mbox{and}\quad Q_q^N=(Q^N_q)_+.
\ee
The $q$-differential operator $Q^N_q$ has the form below and tends
to the differential operator of the $N$-KdV hierarchy as $q$ goes
to 1:
\bea
Q^N_q=D^N_q&+&\frac{\pl}{\pl
t_1}\log\frac{\tau(D^Nc+t)}{\tau(c+t)}D^{N-1}_q\nonumber\\
&+&\left(\sum_{i=0}^{N-1}\frac{\pl^2}{\pl
t_1^2}\log\tau(D^ic+t)\right.\nonumber\\
&-&\sum_{i=0}^{N-2}\lb_{i+1}\left(\frac{\pl}{\pl
t_1}\log\frac{\tau(D^Nc+t)}{\tau(D^{N-1}c+t)}-
\frac{\pl}{\pl
t_1}\log\frac{\tau(D^{i+1}c+t)}{\tau(D^ic+t)}\right)\nonumber\\
&+&\left.\sum_{0\leq i\leq j\leq N-2}\frac{\pl}{\pl
t_1}\log\frac{\tau(D^{i+1}c+t)}{\tau(D^ic+t)}
\frac{\pl}{\pl
t_1}\log\frac{\tau(D^{j+1}c+t)}{\tau(D^jc+t)}\right)D_q^{N-2}+...
\nonumber\\
&\lrg&\left(\frac{\pl}{\pl x}\right)^N+N\frac{\pl^2}{\pl
t^2_1}\log\tau(\bar x+t)\left(\frac{\pl}{\pl x}\right)^{N-2}+...
\eea
\end{corollary}

\proof Note that for $W\in Gr^{(0)}$, $z^NW\subset W$ if and only
if its tau function is of the form $e^{\sum_1^{\iy}
c_{i}t_{iN}}\tau(t)$, with $\pl\tau(t)/ \pl t_{iN}=0$, $i=1,2,...$.
Thus by hypothesis, we have for each
$$
W_k=\span\{\psi_k(t,z),\psi_{k+1}(t,z),...\}
$$
$z^NW_k\subset W_k$ and since $L\psi=z\psi$,
$$
z^N\psi_k=\sum^{N-1}_{j=0}a_j\psi_{k+j}+\psi_{k+N}=(L^N\psi)_k,
$$
and so $L^N$ is upper-triangular, yielding (5.21), which by the
isomorphism $^{\wedge}$ of Lemma 5.3 yields (5.22). From (0.13) and
the relationship between $a_i(y)$ and $b_i(y)$ given in (5.12),
deduce (5.23).

\end{document}